\documentclass[12pt,a4paper, fleqn]{article}

\RequirePackage[l2tabu, orthodox]{nag}

\usepackage{mjheppub}

\usepackage[british]{babel}
\usepackage[utf8]{inputenc}
\usepackage{fancyhdr}
\usepackage{amsmath}
\usepackage{amssymb}
\usepackage{yfonts}
\usepackage[colorinlistoftodos]{todonotes}
\usepackage{color}
\usepackage{mathrsfs}
\usepackage{subcaption}
\usepackage{braket}
\usepackage{xfrac}
\usepackage{dsfont}
\usepackage{amsthm,amssymb,amsmath,epic,eepic,float}
\usepackage{rotating,epsfig,indentfirst,array,varioref}
\usepackage{appendix,marginnote,bbm,tikz,pgf,mathtools}

\usepackage{lipsum}
\usepackage{amsthm,amssymb,amsmath,epic,eepic,float}
\usepackage{rotating,epsfig,indentfirst,array,varioref}
\usepackage{appendix,marginnote,bbm,tikz,pgf,mathtools}

\usepackage{tikz}
\usetikzlibrary{decorations.pathreplacing}
\usetikzlibrary{calc}
\usepackage{pgfplots}
\usepgfplotslibrary{fillbetween}

\pgfplotsset{compat=newest}

\usepackage{empheq} 

\def\leq{\leqslant}
\def\geq{\geqslant}

\newcommand{\bea}{\begin{eqnarray}\displaystyle}
\newcommand{\eea}{\end{eqnarray}}

\catcode`@=12
\relax

\def\ie{\hbox{\it i.e.}}

\def\beq{\begin{equation}}
\def\eeq{\end{equation}}
\def\bea{\begin{eqnarray}}
\def\eea{\end{eqnarray}}
\def\EQ{\begin{equation}}
\def\EN{\end{equation}}

\relax

\usepackage{mathtools}

\numberwithin{figure}{section}
\numberwithin{table}{section}

\usepackage{etoolbox}

\AtBeginEnvironment{figure}{\stepcounter{table}}
\AtBeginEnvironment{table}{\stepcounter{figure}}

\title{\boldmath  Two-point connectivity of two-dimensional critical $Q-$ Potts random clusters on the torus}

\author{Nina Javerzat$^1$, Marco Picco$^2$, Raoul Santachiara$^1$}

\affiliation{$^1$ LPTMS, CNRS (UMR 8626), Univ.Paris-Sud, Universit\'e Paris-Saclay, 91405 Orsay, France}
\affiliation{\vspace{2mm}
$^2$ LPTHE, UMR 7589, Sorbonne Universit\'e and CNRS, France}

\emailAdd{nina.javerzat@u-psud.fr,picco@lpthe.jussieu.fr, raoul.santachiara@u-psud.fr}

\abstract{We consider the two dimensional $Q-$ random-cluster Potts model on the torus and at the critical point. We study the probability for two points to be connected by a cluster for general values of $Q\in [1,4]$. Using a Conformal Field Theory (CFT) approach, we provide the leading topological corrections to the plane limit of this probability. These corrections have universal nature and include, as a special case, the universality class of two-dimensional critical percolation. We compare our predictions to Monte Carlo measurements. Finally, we take Monte Carlo measurements of the torus energy one-point function that we compare to CFT computations. 
}

\begin{document}

\maketitle
\flushbottom

\section{Introduction}
The critical point  of a two-dimensional statistical model can be often characterised in terms of extended objects that, in the continuum limit,  are described by conformal invariant fractal structures \cite{du06}. The study of these fractals provided new insights into the nature of critical phenomena  paving the way to mathematically rigorous approaches \cite{BaBe}. On the one hand, many of the results found so far involve quantities related to two-point correlation functions of  a Conformal Field Theory (CFT). The only exceptions concern observables that satisfy some differential equation and whose definition requires the existence of a boundary, such as crossing probabilities \cite{Caperco92} or SLE interfaces \cite{carsle}. On the other hand, the (bootstrap) solution of a CFT requires the knowledge of three- and four-point correlation functions. Besides some special cases \cite{GaKa,Ka}, the only known bootstrap solutions known to describe statistical critical points are the minimal models. These CFTs have been successful in providing the behaviour  of local observables of critical systems, such as the Ising spin correlation function, but they are too simple to capture the geometry of conformal fractals. The description of these fractals hints therefore at the existence of a CFT whose solution remains an open puzzle. 

The random cluster $Q-$state Potts models \cite{Wupotts} represent an ideal laboratory in this context. This is a one parameter family of models which includes as special cases the spanning forests ($Q\to 0$) \cite{CJSSS04}, the (bond) percolation ($Q=1$), the Ising  ($Q=2$) and the three-state Potts ($Q=3$) spin models, as well as the permutation symmetric point of the Ashkin-Teller model ($Q=4$). For $0\leq Q\leq 4$, the $Q-$state Potts model has a critical point  at which the clusters percolate and have a conformal invariant measure. Natural observables are the cluster connectivities, given by the probability that a number of lattice points belong to the same or different clusters \cite{Deviconn, govi17,govi18}. The conjecture of Delfino and Viti on the three-point connectivities \cite{devi11} has been at the origin of a series of papers \cite{psvd13, dpsv13,ei15,IkJaSa15,clsq18,js18,prs16,prs19} which unveiled  important insights on the still unknown bootstrap solution.

In this paper we focus on the {\it two-point} connectivity {\it on a torus}. This study is motivated by two facts:
\begin{itemize}
\item  In order to increase the number of samplings, Monte Carlo measurements are conveniently taken on doubly periodic lattices \cite{prs16}: a precise knowledge of topological corrections is therefore needed to extract the scaling plane limit which is then compared to the CFT on the sphere predictions.

\item The torus topological effects encode informations on the set of states and on the three-point functions, which are the basic ingredients to solve a CFT.
\end{itemize}

In Section \ref{CFTtorus} we review notions of CFT on a torus and provide the general formulas we will need. In Section \ref{QPottdef} we define the lattice observables and we provide analytical results on their universal finite size behaviours. These results are then compared with the numerical results in Section \ref{MCsec}, where details of the simulations are also discussed. The final conclusions are found in  Section \ref{conclusions}. 
\section{Conformal field theory on a torus}
\label{CFTtorus}
\subsection{Virasoro algebra and its representation}
Consider first a CFT on a plane $z\in (\mathbb{C}\bigcup \{\infty\})$ \cite{rib14} with $T(z)$ and $\bar{T}(\bar{z})$ the  holomorphic and anti-holomorphic component of the stress energy-tensor. The holomorphic stress-energy  modes  $L^{(z)}_{n}$, defined in (\ref{Lndef}) form the Virasoro algebra $\mathcal{V}_c$  with central charge $c$:
\begin{equation}
\label{Vir}
\left[L^{(z)}_{n},L^{(z)}_{m}\right]=(n-m)L^{(z)}_{n+m}+\frac{c}{12}n(n^2-1)\delta_{n,m}.
\end{equation}
The anti-holomorphic modes  $\bar{L}^{(z)}_{n}$ are analogously defined and form a second Virasoro algebra $\overline{\mathcal{V}}_c$, with the same central charge, that commutes with  (\ref{Vir}).

\noindent A highest-weight representation of $\mathcal{V}_c$ is labelled by the conformal dimension $\Delta$:  it contains the primary field $V_{\Delta}$, $L_{n} \ket{V_{\Delta}}=0$ for $n>0$, and its descendants, obtained by acting with the negative modes on the primary state. Given a Young diagram $Y = \{ n_1, n_2\cdots\}$, with $n_{i} \in \mathbb{N}, n_{i}\leq n_{i+1}$,  the fields 
\begin{align}
V_{\Delta}^{(Y)}=L^{(z)}_{-Y} \, V_{\Delta}=L^{(z)}_{-n_1}L^{(z)}_{-n_2}\cdots \;V_{\Delta}\quad (V_{\Delta}^{(\{0\})}=V_{\Delta})
\end{align} 
form a complete basis of the $\Delta$ representation.  The descendant $V^{(Y)}_{\Delta}$ has total dimension $\Delta+|Y|$, where $|Y|=\sum n_i$ is called the level of the descendant. For general $\Delta$, the number of independent descendants is therefore the number of partitions of $|Y|$. The inner product  $H_{\Delta} (Y, Y') $ between descendants is defined as:
\begin{equation}
\label{H}
H_{\Delta} \left( Y, Y' \right) = \lim_{z\to \infty} z^{2\Delta}\;
\left< V_{\Delta}(z) L^{(0)}_{Y}L^{(0)}_{-Y'} V_{\Delta}(0) \right>,
\end{equation} 
and is completely defined by the algebra (\ref{Vir}). For certain values of $\Delta$, see (\ref{paradelta}), the representations are degenerate: they contain a descendant field, usually called the null state, which has vanishing norm. For unitary CFTs, the null state is set to zero. Otherwise, one can have CFTs where null states are not vanishing, like for instance in \cite{RiSa14}. For the sake of simplicity, we will continue to denote the descendant states as $V^{(Y)}_{\Delta}$ even when the presence of a vanishing null state makes their number smaller than the number of partitions. In this case, the notation $Y$ is meant to label the independent non-vanishing descendants.

The spectrum $\mathcal{S}$ of a CFT is formed by the representations of $\mathcal{V}_c \otimes \overline{\mathcal{V}}_c$ appearing in the theory and labelled by the holomorphic and anti-holomorphic dimensions $\Delta,\bar{\Delta}$. In order to simplify the formulas, we use the notations  $(\Delta)_i=\Delta_i,\bar{\Delta}_i$ and  $(\Delta,Y)_i=(\Delta_i,Y_i),(\bar{\Delta}_i,\bar{Y}_i)$. In these notations, a  $\mathcal{V}_c \otimes \overline{\mathcal{V}}_c$ primary field and its descendants are 
\begin{align}
V_{(\Delta)}(z,\bar{z})= V_{\Delta}(z)V_{\bar{\Delta}}(\bar{z}), \quad V_{(\Delta,Y)}(z,\bar{z})= L^{(z)}_{-Y}\bar{L}^{(\bar{z})}_{-\bar{Y}} V_{\Delta}(z)V_{\bar{\Delta}}(\bar{z}).
\end{align}
The product of two primary fields (OPE) can be expanded in terms of the states appearing in the spectrum $\mathcal{S}$ \cite{rib14}: 
\begin{align}
\label{OPE}
V_{(\Delta)_1}(z,\bar{z})\;V_{(\Delta)_2}(0)& \to a^{(\Delta,Y)_3}_{(\Delta)_1,(\Delta)_2}(z,\bar{z}) \;V_{(\Delta,Y)_3}(0),
\end{align}
where the coefficients are factorised as:
\begin{align}
\label{acoeff}
a^{(\Delta,Y)_3}_{(\Delta)_1,(\Delta)_2}(z,\bar{z})= C^{(\Delta)_3}_{(\Delta)_1,(\Delta)_2} \;\beta^{(\Delta_3,Y_3)}_{\Delta_{1},\Delta_2}(z)\;\beta^{(\bar{\Delta},\bar{Y}_3)}_{\bar{\Delta}_{1},\bar{\Delta}_2} (\bar{z}).
\end{align}
One factor is the (model dependent) structure constant  $C^{(\Delta)_3}_{(\Delta)_1,(\Delta)_2}$, the other factor is fixed by the algebra (\ref{Vir}):
\begin{align}
\label{beta}
\beta^{(\Delta_3,Y_3)}_{\Delta_{1},\Delta_2}(z)= z^{-\Delta_1-\Delta_2+\Delta_3+|Y|}\sum_{\substack{Y'\\|Y'|=|Y|}} \; H^{-1}_{\Delta_3}(Y,Y')\;\Gamma_{(\Delta_2,\{0\}),(\Delta_1,\{0\}))}^{(\Delta_3,Y')}, 
\end{align}
where:
\begin{align}
\label{gamma3}
\Gamma_{(\Delta_1,Y_1),(\Delta_2,Y_3)}^{(\Delta_3,Y_3)}=\frac{\left< L^{(\infty)}_{-Y_3}\;V_{\Delta_3}(\infty) L_{-Y_2}^{(1)}V_{\Delta_{2}}(1) L_{-Y_1}^{(0)}  V_{\Delta_{1}}(0)\right>}{\left< V_{\Delta_3}(\infty) V_{\Delta_{2}}(1)   V_{\Delta_{1}}(0)\right>}.
\end{align}
Under the replacement $\Delta_i\to \bar{\Delta}_i$, the above formulas define  $\beta^{(\Delta_3,\bar{Y}_3)}_{\bar{\Delta}_{1},\bar{\Delta}_2} (\bar{z})$ too.
The three-point function  (\ref{gamma3} can be computed in an efficient way by the recursion formulas in \cite{kms10}.

\noindent In the study of the critical random Potts model, the following  series of notations turns out to be very convenient.  The conformal dimension can be parametrised as follows 
\begin{equation}
\label{paradelta}
 \Delta= \Delta_{(r,s)}  = \frac{c-1}{24} + \frac14 \left(r\beta -\frac{s}{\beta}\right)^2.
\end{equation}
A representation is degenerate  if  $r,s \in \mathbb{N^*}$, and has a null state at level $r s$. The symbols 
\begin{align}
V_{\Delta_{(r,s)},\Delta_{(r,s)}}=V_{(r,s)^D},\quad V_{\Delta_{(r,s)},\Delta_{(r,-s)}}=V_{(r,s)}
\end{align}
indicate the diagonal and non-diagonal primary fields and the notations   
\begin{align}
(r,s)^D, \quad (r,s)
\end{align}
 denote the representations associated to $V_{(r,s)^D}$ and $V_{(r,s)}$ respectively. This allows us to use a lighter notations for the structure constants, for instance:
 \begin{align}
  C_{(r_1,s_1),(r_2,s_2)}^{(r,s)^D}=C^{(\Delta_{r_3,s_3},\Delta_{r_3,s_3})}_{\left( \Delta_{r_1,s_1},\Delta_{r_1,-s_1}\right),\left( \Delta_{r_2,s_2},\Delta_{r_2,-s_2}\right)}.
  \end{align}

\noindent A set of these representations is denoted as 
\begin{align}
\mathcal{S}^{D}_{X}=\{(r,s)^D\}_{(r,s)\in X}, \quad  \mathcal{S}_{X}=\{(r,s)\}_{(r,s)\in X},
\end{align}
where $X$ is a given set of pairs $(r,s)$. A third set type is $\mathcal{S}^{\mathrm{quot}}_X$ that contains the degenerate representations with vanishing null state.

\subsection{Torus correlation functions}

So far we have reviewed the properties of a CFT that do not depend on the topology of the surface. The theory of Virasoro algebra on general Riemann surfaces can be found in \cite{EgOo87}. Let us consider now a CFT on a torus with periods $\omega_1$ and $\omega_2$. In the numerical simulations one usually considers doubly periodic rectangular lattices of size $M\times N$, where $M,N \in \mathbb{R}_{>0}$. We therefore set:
\begin{equation}
\label{toruspar}
\omega_1=i M,\quad \omega_2=N, \quad \tau=\frac{\omega_{1}}{\omega_{2}}=i \frac{M}{N}, \quad q=e^{2 \pi i \tau}.
\end{equation}
The results we will obtain can be of course generalized to the case $\mathrm{Re}\;\tau\neq 0$. In the following,  we represent the torus as a finite cylinder of size $N$ with the ends, at distance $M=O(N)$, glued together. Accordingly, we use the map 
\begin{equation}
\label{map}
w=-i\; \frac{N}{2 \pi}  \ln z
\end{equation}
 sending the plane ($z$) to an infinite cylinder ($w$) of size  $N$.  
 
\noindent We define a general field $V^{\mathcal{C},N}_{(\Delta,Y)}$ on the cylinder of size $N$ as:
\begin{align}
\label{Lcdef}
V^{\mathcal{C},N}_{(\Delta,Y)}(w,\bar{w})=L^{\mathcal{C},(w)}_{-Y}\bar{L}^{\mathcal{C},(\bar{w})}_{-\bar{Y}}\; V^{\mathcal{C},N}_{(\Delta)}(w,\bar{w}), 
\end{align}
where  $L_{-n}^{\mathcal{C},(w)}$, $\bar{L}_{-n}^{\mathcal{C},(w)}$ are the conformal generators on the cylinder. 
They are related to the fields on the cylinder of size $N=1$ by a factor arising from their transformation under (\ref{map}), see Appendix \ref{torusdet}:
\begin{equation}
V^{\mathcal{C},N}_{(\Delta,Y)}(w,\bar{w}) = \left(\frac{2\pi }{N}\right)^{\Delta+\bar{\Delta}+|Y|+|\bar{Y}|}V^{\mathcal{C},N=1}_{(\Delta,Y)}(w,\bar{w})
\end{equation}
Henceforth,  we will often omit the symbol $\mathcal{C},N$ when the field on the cylinder is a primary, i.e.  $V^{\mathcal{C},N}_{(\Delta)}(w)\to V_{(\Delta)}(w)$. The relation between  $L_{-n}^{\mathcal{C},(w)}$ and $L_{-n}^{(z)}$ is obtained using the transformation of $T$ under the map (\ref{map}) see  Appendix \ref{torusdet}.

\noindent The torus function $\left< \prod_i V_{(\Delta_i)}\right>_{\tau}$ corresponds to the trace of the transfer matrix with field insertions. The one-point torus correlation function can be associated with the following diagram:
\begin{center}
\begin{tikzpicture}[baseline=(current  bounding  box.center), very thick, scale = .4]
\draw (0,2) node [above] {$V_{(\Delta)}$} -- (0,-1);
\draw (0,-3) circle(1.9cm);
\draw (0,-5) node [below] {$\mathcal{S}_{\mathrm{int}}$};
\end{tikzpicture}
\end{center}
where $\mathcal{S}_{\mathrm{int}}\in \mathcal{S} $ is the set of representations that propagate along the $M$ direction and  whose fusion with themselves contains the representation $(\Delta)$. It takes the form:
\begin{align}
\label{1ptorus}
\langle V_{(\Delta)}\rangle_{\tau}& = \frac{1}{Z}\text{Tr}_{\mathcal{S}_{\mathrm{int}}} \left(q^{L_{0}^{\mathcal{C},(\infty)}}\bar{q}^{\bar{L}_{0}^{\mathcal{C},(\infty)}} V_{(\Delta)}(0)\right)=\frac{1}{Z}\sum_{(\Delta_{\mathrm{int}})\in \mathcal{S}_{\mathrm{int}}}  C^{(\Delta)}_{(\Delta_{\mathrm{int}}),(\Delta_{\mathrm{int}})} \mathcal{F}_{\Delta}^{\Delta_{\mathrm{int}}}(q)\;\mathcal{F}_{\bar{\Delta}}^{\bar{\Delta}_{\mathrm{int}}}(\bar{q}),
\end{align}
where $L_{0}^{\mathcal{C},(\infty)}=L_{0}^{(0)}-\frac{c}{24}$ and 
$\mathcal{F}_{\Delta}^{\Delta_{\mathrm{int}}}(q)$ is the torus conformal block:
\begin{multline}
\label{defF}
\mathcal{F}_{\Delta}^{\Delta_{\mathrm{int}}}(q)=q^{\Delta_{\mathrm{int}}-\frac{c}{24}}\; \bar{q}^{\bar{\Delta}_{\mathrm{int}}-\frac{c}{24}}\;\sum_{\substack{Y,Y'\\|Y|=|Y'|}}\; q^{|Y|} H^{-1}_{\Delta_{\mathrm{int}}}(Y,Y') \;\;\Gamma_{(\Delta_{\mathrm{int}},Y),(\Delta,\{\emptyset\})}^{(\Delta_{\mathrm{int}},Y')}\\
= q^{\Delta_{\mathrm{int}}-\frac{c}{24}}\; \bar{q}^{\bar{\Delta}_{\mathrm{int}}-\frac{c}{24}}\;\left(1+\frac{2\Delta_{\mathrm{int}}+\Delta(\Delta-1)}{2\Delta_{\mathrm{int}}}q+\cdots\right)
\end{multline}
see (\ref{H})-(\ref{gamma3}). The torus partition function $Z$ can be related to the identity one-point function, for which $\mathcal{S}_{\mathrm{int}}= \mathcal{S}$, 
\begin{align}
Z = \text{Tr}_{\mathcal{S}} \left(q^{L_{0}^{\mathcal{C},(\infty)}}\bar{q}^{\bar{L}_{0}^{\mathcal{C},(\infty)}}\right)=\sum_{(\Delta_{\mathrm{int}})\in \mathcal{S}} \; \mathcal{F}_{0}^{\Delta_{\mathrm{int}}}(q)\;\mathcal{F}_{0}^{\bar{\Delta}_{\mathrm{int}}}(\bar{q}).
\end{align}
The computation of $\mathcal{F}_{0}^{\Delta_{\mathrm{int}}}(q)$ using  recursion relations is discussed  in \cite{jsf18}.

\noindent The $s-$ channel expansion of the torus two-point function $\left< V_{(\Delta)_1}(w)V_{(\Delta)_2}(0)\right>_{\tau}$ is described by the diagram:
\begin{center}
\begin{tikzpicture}[baseline=(current  bounding  box.center), very thick, scale = .4]
\draw (-3,3) node [left] {$V_{(\Delta)_1}(w)$}--(0,0)--(3,3) node[right]{$V_{(\Delta)_2}(0)$};
\draw (0,0)  -- (0,-3);
\draw (0,-1.5) node[right]{$ \mathcal{S}_{\mathrm{int}}$};
\draw (0,-5) circle(1.9cm);
\draw (0,-7) node [below] {$\mathcal{S}_{\mathrm{int}_2}$};
\end{tikzpicture}
\end{center}
where $\mathcal{S}_{\mathrm{int}}$ contains the fields appearing in the fusion $V_{(\Delta)_1}V_{(\Delta)_2}$ and $\mathcal{S}_{\mathrm{int}_2}$  is the spectrum of the one-point torus function of the fields in $\mathcal{S}_{\mathrm{int}}$.
One can show, see Appendix \ref{torusdet},  that the two-point torus function can be expanded as:
\begin{align}
\label{2ptorus}
\langle V_{(\Delta)_1}(w) V_{(\Delta)_2}(0)\rangle_{\tau}& =\left(\frac{N}{2\pi}\right)^{-\Delta_1-\Delta_2-\bar{\Delta}_1-\bar{\Delta}_2}\sum_{ (\Delta,Y)_{\mathrm{int}}\in \mathcal{S}_{\mathrm{int}}}  a^{(\Delta,Y)_{\mathrm{int}}}_{(\Delta)_1,(\Delta)_2}\left(\frac{2\pi  w}{N}\right) \left< V^{\mathcal{C}}_{(\Delta,Y)}\right>_{\tau}.
\end{align}

\section{$Q-$Potts random cluster model}
\label{QPottdef}
  
Let us consider a rectangular lattice $N\times M$ with periodic boundary conditions in the two directions.  The  edges of the graph carry a bond with probability $p$, or no bond with probability $1-p$. According to these bonds, the lattice is split into a disjoint union of connected clusters. The random cluster $Q$-state Potts model \cite{FK72}  is defined by the partition function 
\begin{align}
 \mathcal{Z}_Q=\sum_{\mathcal{G}} Q^{\#\,\mathrm{clusters}} p^{\#\, \mathrm{bonds}} (1-p)^{\#\, \text{edges without bond}}.
 \label{eq:z}
\end{align}

\noindent At the critical value
\begin{align}
\label{criticalp}
p=p_c=\frac{\sqrt{Q}}{\sqrt{Q}+1},
\end{align}
the probability that there exists a percolating cluster jumps from $0$ to $1$, in the limit of infinite lattice size. The model becomes conformally invariant in the scaling limit, and has central charge $c$:
\begin{align}
c = 1-6\left(\beta -\beta^{-1}\right)^2, \qquad Q = 4\cos^2\pi \beta^2 \quad \text{with} \quad \tfrac12 \leq  \beta^2 \leq 1\ .
\label{cq}
\end{align} 

\noindent The scaling limit $Z_Q$ of the  Potts partition function (\ref{eq:z}) at the critical point (\ref{criticalp}) was computed in \cite{FraSaZu87}:
\begin{equation}
Z_Q = \mathrm{Eq.}\;\mathrm{(4.8)}\quad  \mathrm{of}\; \;$\cite{FraSaZu87}$, \quad \mathrm{with}\;e_0\to2-2\beta^2,\quad  g\to4\beta^2,  \quad  h_{s,r}\to \Delta_{(-2 r, \frac{s}{2})}
\end{equation}
\noindent The corresponding total spectrum is:  
\begin{equation}
\label{sppotts}
\mathcal{S}^{\mathrm{Potts}}=\mathcal{S}^{D,\mathrm{quot}}_{(1,\mathbb{N^*})}\bigcup_{\substack{j\geq 2 \\ M | j, p\wedge M=1}} \mathcal{S}_{(j,\mathbb{Z}+\frac{p}{M})} \bigcup \;\mathcal{S}_{(0,\mathbb{Z}+\frac12)} .
\end{equation}
The multiplicities associated to the above sectors have also been computed \cite{FraSaZu87} and, for general $Q$, assume general real values. We refer the reader to \cite{js18} for a derivation  of (\ref{sppotts}) from the representations of Temperley-Lieb type algebras.
\noindent $\mathcal{S}^{D,\mathrm{quot}}_{(1,\mathbb{N^*})}$ is the  thermal sector \cite{dofa_pl85} and  contains the identity  and the energy field:
$$
\mathrm{Identity \;field}= V_{(1,1)^D},\quad \mathrm{Energy\;field} =V_{(1,2)^D}.
$$
\noindent The space of $n-$point cluster connectivities has been  defined in \cite{Deviconn}. Here we will focus only on the two-point connectivities:  
\begin{align}
\quad p_{12} & =\text{Probability}(w_1,w_2\text{ are in the same cluster}).
\label{2pointdef}
\end{align}
At  the critical point (\ref{criticalp}) and in the plane limit $N,M\to \infty$, the Coulomb gas approach \cite{henkel2012conformal} determines  the scaling limit of the probability  $p_{12}$ :  
\begin{equation}
\label{scalingp120}
\mathrm{Plane\; scaling \;limit:\;  }\;p_{12} = c_0\; |w|^{-4\Delta_{(0,\frac12)}}\;,\quad w=w_1-w_2,
\end{equation}
where $c_0$ is a non-universal constant, see Section \ref{MCsec}. From the above equation one sees that, in the plane, the two-point connectivity is related to the plane two-point function of  the \begin{align}
\mathrm{Connectivity \;field}= V_{(0,\frac12)},
\end{align}
belonging to the magnetic sector $\mathcal{S}_{(0,\mathbb{Z}+\frac{1}{2})}$  \cite{saleur87}. It is natural to assume that the relation between  $p_{12}$ and the $V_{(0,\frac12)}$ two-point function holds on the torus, i.e.:
\begin{equation}
\label{scalingp12}
\mathrm{Torus \;scaling \;limit:}\;p_{12} = c_0 \;\left< V_{(0,\frac12)}(w)V_{(0,\frac12)}(0)\right>_{\tau},\quad w=w_1-w_2.
\end{equation}
Let us mention that a rigorous proof of (\ref{criticalp}) has been obtained recently in \cite{beffaracopin10} where the behaviour of the probability (\ref{2pointdef}) in the sub-critical regime $p<p_c$ and on the torus  was also studied.

\section{Two-point Potts torus connectivity}
According to Monte Carlo simulations (see Section \ref{MCsec}) while $(0,\frac12)$ is the field in (\ref{sppotts}) with the smallest non-zero conformal dimension, the leading topological correction is given by the energy  state $(1,2)^D$. The contribution from the second thermal operator $(1,3)^D$ is also visible at  $Q\sim 3$. Based on these observations, we assume that $\left<V_{(0,\frac12)}(w)V_{(0,\frac12)}(0)\right>_{\tau}$ is given by (\ref{2ptorus}) with $\mathcal{S}_{\mathrm{int}}=\mathcal{S}^{D,\mathrm{quot}}_{(1,\mathbb{N}^*)}$. In particular we compute the contributions of the first three dominant channels:  
\begin{equation}
\label{asspect}
\mathcal{S}_{\mathrm{int}}=\{(1,1)^D,(1,2)^D, (1,3)^D\}$$
\end{equation} 
The agreement between Monte Carlo and analytic results presented below confirms that this truncated spectrum (\ref{asspect}) provides a good approximation to $\left<V_{(0,\frac12)}(w)V_{(0,\frac12)}(0)\right>_{\tau}$. Some arguments going in this direction come also  from the analysis in \cite{js18,prs19} where the spectrum of all independent four-point connectivities has been determined. In particular, it was shown that the asymptotic of the probability $p_{12}\cap p_{34}$ (related to $P_0+P_1$ in \cite{prs19}), in the limit $z_2-z_1\to 0$ and $z_3-z_2>>1$, is dominated by the low lying states $(1,1)^D, (1,2)^D,(1,3)^D,(2,0),\cdots$. In this limit one expects that $p_{12}\cap p_{34}\sim \;p_{12} \;p_{34}+\mathrm{corrections}$, where the corrections are produced by the configurations  which correlate the $p_{12}$ and $p_{13}$ probabilities and which are associated to the state $(2,0)$\cite{prs19}. 

\noindent
In the limit:
\begin{align}
\label{limto}
N\to \infty,\quad  \frac{M}{N}\to O(1),\quad 1<<w<< N,
\end{align}
using the expression for the two-point function (\ref{2ptorus}) with the internal spectrum (\ref{asspect}) we obtain the following $N^{-1}$ expansion
\begin{align}
\label{2ptconnto}
&\left< V_{(0,\frac{1}{2})}(w) V_{(0,\frac12)}(0)\right>_{\tau}= \left|w\right|^{-4\Delta_{(0,\frac12)}}\sum_{X \in \{(1,1)^D,(1,2)^D,(1,3)^D\}} C^{X}_{(0,\frac12),(0,\frac12)}\;\left|\frac{2\pi w}{N}\right|^{2\Delta_{X}}\left(\left< V_{X}\right>_{\tau}+\right.\nonumber \\
&  + \left(\frac{2\pi}{N}\right)^2\beta^{\{2\}}_{X}\left( w^2\;\left< L^{\mathcal{C},(0)}_{-2}V_{X} \right>_{\tau}+\bar{w}^2\;\left< \bar{L}^{\mathcal{C},(0)}_{-2}V_{X} \right>_{\tau}\right)+\nonumber \\
&+\left.\left(\frac{2\pi}{N}\right)^3\beta^{\{3\}}_{X}\left( w^3\;\left< L^{\mathcal{C},(0)}_{-3}V_{X} \right>_{\tau}+\bar{w}^3\;\left< \bar{L}^{\mathcal{C},(0)}_{-3}V_{X} \right>_{\tau}\right)+\cdots\right),
\end{align}
where we set $\beta^{Y}_{X}=\beta^{(\Delta_X,Y)}_{\Delta_{(0,\frac12)},\Delta_{(0,\frac12)}}$. Note that descendants of the type $L^{\mathcal{C}}_{-1}L^{\mathcal{C}}_{-Y} V_{X}$ are total derivatives and their torus one-point functions vanish. 

\noindent The main message here is that the leading topological correction for the two-point connectivity is given, for $1\leq Q\leq 4$ by the energy $(1,2)^D$ state. Given two-points $w_1, w_2$ on a torus  (\ref{toruspar}) and at distance $r=|w_1-w_2|$, the scaling limit of the probability (\ref{2pointdef}) is:
\begin{align}
\label{Qgen}
p_{12}\;=\frac{c_0}{r^{4\Delta_{(0,\frac12)}}}\Bigg[&1+ \left(\frac{r}{N}\right)^{2\Delta_{(1,2)}}\left( \frac{(2\pi)^{2\Delta_{(1,2)}}}{Z_Q(q)}(Q-1)\;\left[C^{(1,2)^D}_{(0,\frac12),(0,\frac12)}\right]^2\;q^{2\left(\Delta_{(0,\frac12)}-\frac{c}{24}\right)}\left(1+O(q)\right)\right)\Bigg.\nonumber \\
&\Bigg.+O\left(\left(\frac{r}{N}\right)^2\right)\Bigg]
\end{align}
where $c_0$ is a non-universal constant evaluated in Table (\ref{Table2pt}), and $C^{(1,2)^D}_{(0,1/2),(0,1/2)}$ is given in (\ref{ensc}). At the critical percolation $Q=1$ point, we have:
\begin{align}
\label{critperco}
p_{12}=\frac{c_0}{r^{\frac{5}{24}}}\Bigg[&1+\left(\frac{r}{N}\right)^{\frac{5}{4}}\left((2\pi)^{\frac{5}{4}}\pi\sqrt{3}\left(\frac{4}{9}\frac{\Gamma(\frac74)}{\Gamma(\frac14)}\right)^2\; e^{-\frac{5 \pi}{24} \frac{M}{N}}+O\left(e^{-\frac{53}{24} \pi \frac{M}{N}}\right)\right)+\Bigg.\nonumber \\ \Bigg.&+O\left(\left(\frac{r}{N}\right)^2\right)\Bigg].
\end{align}
\noindent The formula (\ref{Qgen}) represents, at the best of our knowledge, a new analytic result on the universal properties of general $Q$ random Potts critical clusters and, in particular, of the critical percolation clusters (\ref{critperco}). The derivation of (\ref{Qgen}) and (\ref{critperco}), of the next $\frac{r}{N}$ sub-leading topological terms and of the systematic computation of the $q$ expansion, are given below.

\subsection{Identity channel contributions}
\label{idesec}

The leading contribution to (\ref{2ptconnto}) comes from the identity. In particular we have:
\begin{subequations}
\begin{align}
\mathrm{Leading\; :}&\;\left|w\right|^{-4\Delta_{(0,\frac12)}} \quad \left(\mathrm{plane \;limit}\right), \\
\mathrm{Sub-leading\;:}&\;\left|w\right|^{-4\Delta_{(0,\frac12)}}\left[\;\left(\frac{w}{N}\right)^2 \;c_{T}+ \left(\frac{\bar{w}}{N}\right)^2 \;c_{\bar{T}}\right]\\
\mathrm{Next \;to\; sub-leading:}&\; O\left(\frac{1}{N^4}\right)
\end{align}
\end{subequations}
The dominant term corresponds to the plane limit while the sub-leading factors $c_T$ and $c_{\bar{T}}$:

\begin{align}
\label{cT}
c_{T}=  \frac{2\Delta_{(0,\frac12)}}{c}\;\left< T^{\mathcal{C}}\right>_{\tau},\quad c_{\bar{T}}= \frac{2\Delta_{(0,\frac12)}}{c}\;\left< \bar{T}^{\mathcal{C}}\right>_{\tau}
\end{align} 

are proportional to the stress energy one-point function, with 
\begin{equation}
\langle T^{\mathcal{C}}\rangle_{\tau}= i\pi\partial_\tau \;\text{log}\;Z_Q.
\end{equation}
In Fig. \ref{fig:T} below, we plot $c_T$ as a function of $Q$ and for different $\tau$, i.e. for different ratios $\frac{M}{N}$:
\begin{figure}
\centering
\begin{tikzpicture}
\begin{axis}[
	legend cell align=center,
	xlabel={$Q$},
	ylabel={$c_T$},
	xmin=1,
  xmax=4,
	legend pos=outer north east]
	]

\addlegendentry{$\frac{M}{N} = \infty$}    
\addplot[red, smooth, mark= none] 
coordinates{
(1.001, 0.171399)(1.051, 0.173931)(1.101, 0.176341)(1.151, 
0.178639)(1.201, 0.180832)(1.251, 0.182927)(1.301, 0.18493)(1.351, 
0.186846)(1.401, 0.188682)(1.451, 0.19044)(1.501, 0.192126)(1.551, 
0.193742)(1.601, 0.195293)(1.651, 0.19678)(1.701, 0.198208)(1.751, 
0.199578)(1.801, 0.200892)(1.851, 0.202154)(1.901, 0.203364)(1.951,
0.204525)(2.001, 0.205639)(2.051, 0.206706)(2.101, 0.207728)(2.151,
0.208706)(2.201, 0.209642)(2.251, 0.210536)(2.301, 0.21139)(2.351, 
0.212203)(2.401, 0.212978)(2.451, 0.213713)(2.501, 0.214411)(2.551,
0.215071)(2.601, 0.215694)(2.651, 0.216279)(2.701, 0.216828)(2.751,
0.217339)(2.801, 0.217813)(2.851, 0.21825)(2.901, 0.218649)(2.951, 
0.219009)(3.001, 0.219331)(3.051, 0.219612)(3.101, 0.219852)(3.151,
0.22005)(3.201, 0.220203)(3.251, 0.220311)(3.301, 0.220369)(3.351, 
0.220376)(3.401, 0.220326)(3.451, 0.220217)(3.501, 0.220042)(3.551,
0.219792)(3.601, 0.219459)(3.651, 0.21903)(3.701, 0.218485)(3.751, 
0.2178)(3.801, 0.216933)(3.851, 0.215816)(3.901, 0.214314)(3.951, 
0.212079)
};

\addlegendentry{$\frac{M}{N} = 5$}    
\addplot[green, smooth, mark= none] 
coordinates{
(1.001, 0.159617)(1.051, 0.162153)(1.101, 0.164569)(1.151, 
0.166871)(1.201, 0.169067)(1.251, 0.171164)(1.301, 0.173169)(1.351, 
0.175085)(1.401, 0.176919)(1.451, 0.178674)(1.501, 0.180355)(1.551, 
0.181965)(1.601, 0.183507)(1.651, 0.184985)(1.701, 0.1864)(1.751, 
0.187757)(1.801, 0.189056)(1.851, 0.1903)(1.901, 0.19149)(1.951, 
0.19263)(2.001, 0.193719)(2.051, 0.19476)(2.101, 0.195754)(2.151, 
0.196703)(2.201, 0.197606)(2.251, 0.198465)(2.301, 0.199282)(2.351, 
0.200056)(2.401, 0.200789)(2.451, 0.20148)(2.501, 0.202131)(2.551, 
0.202741)(2.601, 0.203312)(2.651, 0.203842)(2.701, 0.204332)(2.751, 
0.204783)(2.801, 0.205192)(2.851, 0.205561)(2.901, 0.205889)(2.951, 
0.206174)(3.001, 0.206417)(3.051, 0.206615)(3.101, 0.206768)(3.151, 
0.206874)(3.201, 0.20693)(3.251, 0.206935)(3.301, 0.206885)(3.351, 
0.206776)(3.401, 0.206605)(3.451, 0.206366)(3.501, 0.206052)(3.551, 
0.205653)(3.601, 0.20516)(3.651, 0.204556)(3.701, 0.20382)(3.751, 
0.202923)(3.801, 0.201818)(3.851, 0.200425)(3.901, 0.198591)(3.951, 
0.195915)
};

\addlegendentry{$\frac{M}{N} = 2.5$}    
\addplot[gray, smooth, mark= none] 
coordinates{
(1.001, 0.110865)(1.051, 0.112415)(1.101, 0.113877)(1.151, 
0.115257)(1.201, 0.11656)(1.251, 0.11779)(1.301, 0.118953)(1.351, 
0.120051)(1.401, 0.121089)(1.451, 0.122069)(1.501, 0.122994)(1.551, 
0.123867)(1.601, 0.12469)(1.651, 0.125465)(1.701, 0.126195)(1.751, 
0.12688)(1.801, 0.127523)(1.851, 0.128125)(1.901, 0.128687)(1.951, 
0.129211)(2.001, 0.129698)(2.051, 0.130148)(2.101, 0.130564)(2.151, 
0.130945)(2.201, 0.131292)(2.251, 0.131607)(2.301, 0.131889)(2.351, 
0.13214)(2.401, 0.13236)(2.451, 0.132548)(2.501, 0.132707)(2.551, 
0.132834)(2.601, 0.132932)(2.651, 0.133)(2.701, 0.133038)(2.751, 
0.133046)(2.801, 0.133024)(2.851, 0.132972)(2.901, 0.132889)(2.951, 
0.132775)(3.001, 0.132629)(3.051, 0.132451)(3.101, 0.132239)(3.151, 
0.131993)(3.201, 0.131711)(3.251, 0.131391)(3.301, 0.131032)(3.351, 
0.130632)(3.401, 0.130186)(3.451, 0.129692)(3.501, 0.129145)(3.551, 
0.12854)(3.601, 0.127868)(3.651, 0.127119)(3.701, 0.12628)(3.751, 
0.125331)(3.801, 0.124241)(3.851, 0.122957)(3.901, 0.121377)(3.951, 
0.119238)
};

\addlegendentry{$\frac{M}{N} = 1$}    
\addplot[blue, smooth, mark= none] 
coordinates{
(1.001, -0.000000000000118748)(1.051, -0.00000000000000461893)(1.101, 
-0.00000000000000154259)(1.151, -0.00000000000000102384)(1.201, 
-0.00000000000000114505)(1.251, -0.000000000000000455099)(1.301, 
-0.000000000000000627941)(1.351, -0.000000000000000641619)(1.401, 
-0.000000000000000743624)(1.451, -0.000000000000000328336)(1.501, 
-0.000000000000000366979)(1.551, -0.000000000000000662942)(1.601, 
-0.00000000000000024153)(1.651, -0.000000000000000609247)(1.701, 
-0.000000000000000459976)(1.751, -0.00000000000000056887)(1.801, 
-0.00000000000000035336)(1.851, -0.000000000000000247911)(1.901, 
-0.000000000000000193935)(1.951, -0.000000000000000255674)(2.001, 
-0.000000000000000379405)(2.051, -0.000000000000000163269)(2.101, 
-0.000000000000000309852)(2.151, -0.00000000000000035356)(2.201, 
-0.000000000000000224559)(2.251, -0.0000000000000000535782)(2.301,
-0.000000000000000614592)(2.351, -0.000000000000000196123)(2.401, 
-0.000000000000000329021)(2.451, -0.000000000000000360931)(2.501, 
-0.00000000000000013007)(2.551, -0.000000000000000250267)(2.601, 
-0.000000000000000120507)(2.651, -0.000000000000000232323)(2.701, 
0.)(2.751, -0.000000000000000252503)(2.801,
-0.000000000000000313716)(2.851, -0.000000000000000235955)(2.901, 
-0.000000000000000391416)(2.951, -0.0000000000000000947488)(3.001,
-0.0000000000000000917947)(3.051, -0.0000000000000001483)(3.101, 
0.)(3.151, -0.0000000000000000837251)(3.201, 
-0.000000000000000189621)(3.251, -0.0000000000000000526049)(3.301,
-0.000000000000000204377)(3.351, -0.0000000000000000496402)(3.401,
-0.000000000000000217071)(3.451, -0.0000000000000000703243)(3.501,
-0.0000000000000000911404)(3.551, -0.000000000000000155033)(3.601,
-0.000000000000000258318)(3.651, 0.0000000000000000627569)(3.701, 
-0.0000000000000000406435)(3.751, -0.000000000000000078923)(3.801,
0.0000000000000000382806)(3.851, -0.00000000000000012979)(3.901, 
-0.000000000000000107505)(3.951, -0.0000000000000000172253)
};
\end{axis}
\end{tikzpicture}
\caption{ }
\label{fig:T}
\end{figure}
\noindent For a square torus, $M=N$ and $\langle T^{\mathcal{C}}\rangle_{\tau} =\langle \bar{T}^{\mathcal{C}}\rangle_{\tau} = 0$, for all $Q$. This is the reason the $N^{-2}$ corrections were not visible in the fits in (\cite{prs19}). In the cylinder limit $M/N\to \infty$ one recovers the well known result $\langle T^{\mathcal{C}}\rangle_{i\infty}=(2\pi)^2\frac{c}{24}$.
It is interesting to stress that the $\lim_{c\to 0}\frac{2\Delta_{(0,\frac12)}}{c}\left< T^{\mathcal{C}}\right>_{\tau}$ is finite and different from zero. No subtleties, arising from the existence at $c=0$ of a logarithmic partner of the stress energy tensor, seem to emerge. Indeed one can write
\begin{equation}
Z_Q = 1+O(Q-1),
\end{equation}
as one can see by putting $Q=1$ in (\ref{eq:z}), which gives a finite limit for $c_T$.

The  next corrections from the identity channel appear at order $N^{-4}$ and are related to the propagation of the identity descendants  $\langle T^{\mathcal{C}}\bar{T}^{\mathcal{C}}\rangle_{\tau}$, $\langle L^{\mathcal{C}}_{-4} \mathrm{Id}\rangle_{\tau}$ and $\langle \bar{L}^{\mathcal{C}}_{-4} \mathrm{Id}\rangle_{\tau}$.   

\subsection{Energy channel contributions}
\label{energycontr}
Besides the identity, the energy  $V_{(1,2)^D}$ field has the lowest dimension in $\mathcal{S}^{D,\mathrm{quot}}_{(1,\mathbb{N}^*)}$. The $(1,2)^D$ contribution to (\ref{2ptconnto}) is given by
\begin{subequations}
\begin{align}
\label{ener}
&\mathrm{Leading :}\left|w\right|^{-4\Delta_{(0,\frac12)}}\left(\frac{|w|}{N}\right)^{2\Delta_{(1,2)}} c_{(1,2)},\\
&\mathrm{Sub-leading:}\;O\left(\frac{1}{N^{2\Delta_{(1,2)}+4}}\right)
\end{align}
\end{subequations}
where:
\begin{align}
\label{c12}
c_{(1,2)}=\left(2\pi\right)^{2\Delta_{(1,2)}}\;C^{(1,2)^D}_{(0,\frac12),(0,\frac12)}  \left< V_{(1,2)^D}\right>_{\tau}
\end{align}
\noindent We can compute the  one-point function  $\left< V_{(1,2)^D}\right>_{\tau}$ by using the vanishing of the $(1,2)^D$ null state, which determines the OPE \cite{rib14}:
\begin{equation}
V_{(1,2)} \times V_{(r,s)}  \to V_{(r,s+1)}  \oplus V_{(r,s-1)} 
\end{equation}
$(0,\frac12)$ is the only representation which satisfies both the above OPE and
\begin{equation}
V_{(1,2)} \times V_{(0,\frac12)}  \to V_{(0,\frac12)}.
\end{equation}
Therefore the one-point function  $\left< V_{(1,2)^D}\right>_{\tau}$ gets contribution only from the propagation of the $(0,\frac12)$ state, i.e.  $\mathcal{S}_{\mathrm{int}}=\{(0,\frac12)\}$ in (\ref{1ptorus}). This property was pointed out in \cite{zuber90} where the energy one-point function for minimal models was computed in terms of a Coulomb gas integral. Collecting all these facts, we obtain:
\begin{align}\label{e1pt}
\langle V_{(1,2)^D}\rangle_{\tau} = &\frac{Q-1}{Z_Q}\;C^{(1,2)^D}_{(0,\frac12),(0,\frac12)}\left|\mathcal{F}^{\Delta_{(0,\frac12)}}_{\Delta_{(1,2)}}(q)\right|^2\nonumber \\
=& \frac{Q-1}{Z_Q}\;C^{(1,2)^D}_{(0,\frac12),(0,\frac12)} \;|q|^{2\left(\Delta_{(0,\frac12)}-\frac{c}{24}\right)}\left|1+\frac{2\Delta_{(0,\frac12)}+\Delta_{(1,2)}(\Delta_{(1,2)}-1)}{2\Delta_{(0,\frac12)}}q+\cdots\right|^2
\end{align}
where the factor $Q-1$ comes from the multiplicity of the $\mathcal{S}_{(0,\mathbb{Z}+\frac12)}$ sector computed in \cite{FraSaZu87} and the structure constant is given by:
\begin{align}
\label{ensc}
C^{(1,2)^D}_{(0,\frac12),(0,\frac12)} &= \beta^4\frac{\gamma\left(-\frac{1}{2}\right)}{\gamma\left(-\frac{1}{2\beta^2}\right)}\sqrt{\gamma\left(\frac{1}{\beta^2}\right)\gamma\left(2-\frac{2}{\beta^2}\right)}, \quad \gamma(x)=\frac{\Gamma(x)}{\Gamma(1-x)}
\end{align}

\noindent 
The next energy contributions come from the descendants $ L_{-2}^{\mathcal{C}} V_{(1,2)^D}$ and $\bar{L}_{-2}^{\mathcal{C}} V_{(1,2)^D}$. The null state in the representation $(1,2)^D$ is 
\begin{equation}
\chi = \left(-\beta^2 (L^{(1)}_{-1})^2+L^{(1)}_{-2}\right)V_{(1,2)^D}(1).
\end{equation}
Using (\ref{cgen}) 
\begin{equation}
L_{-2}^{\mathcal{C},(0)} = \left(\frac{2\pi i}{N}\right)^2 \left(L_{-2}^{(1)}-\frac{c}{24}-\frac{13}{12} L^{(1)}_0\right)
\end{equation}
and setting the null vector to zero 
\begin{equation}
L^{(1)}_{-2} V_{(1,2)^D}(1)=\beta^2 (L^{(1)}_{-1})^2\;V_{(1,2)^D}(1)
\end{equation}
leads to 
\begin{equation}
\left< L_{-2}^{\mathcal{C},(0)} V_{(1,2)^D}(0)\right>_{\tau}=\left< \bar{L}_{-2}^{\mathcal{C},(0)} V_{(1,2)^D}(0)\right>_{\tau}=0,
\end{equation}
which explains why the sub-leading corrections in (\ref{ener}) are found in the fourth level descendants of the energy (the third level descendant is a total derivative).
Using the expression of the one-point function (\ref{e1pt}) in (\ref{ener}) with $r=|w|$, one obtains our result (\ref{Qgen}).

\noindent At the critical percolation point $Q=1$, the bond probabilities, associated to the energy field (see next section), are independent. The CFT energy one-point function (\ref{e1pt}), which actually probes the fluctuation induced corrections to the bulk constant value, vanishes at $Q=1$. On the other hand, the vanishing of the one-point function is exactly cancelled by the divergence in the structure constant (\ref{ensc}), thus providing a non-zero contribution to $\lim_{Q\to 1} C^{(1,2)^D}_{(0,1/2),(0,1/2)}\langle V_{(1,2)^D}\rangle_{\tau}$.  The result is given in (\ref{critperco}).\par

\noindent When $M\neq N$, we have seen that we have a $N^{-2}$ contribution of the energy tensor to the topological corrections. Even if this term is sub-leading in the  parameter $\frac{r}{N}$, $r=|w|$, in finite size simulations it can interfere or even be dominant with respect to the energy contribution. In Fig. \ref{fig:TE} we plot as function of $Q$, and for different ratios $\frac{M}{N}$, the regimes of $\frac{r}{N}$ dominated by the energy (below the curve) or by the stress-energy (above the curve) topological corrections:

\begin{figure}
\centering
\begin{tikzpicture}
\begin{axis}[
	legend cell align=center,
	xlabel={$Q$},
	ylabel={$\frac{r}{N}$},
	ymin = 0,
	xmin=1,
  xmax=4,
	legend pos=outer north east]
	]

\addlegendentry{$\frac{M}{N} = 5$}    
\addplot[name path = f, blue, smooth, mark= none] 
coordinates{
(1.01, 0.0142422)(1.06, 0.0148335)(1.11, 0.0153586)(1.16, 
0.0158817)(1.21, 0.0164035)(1.26, 0.0169247)(1.31, 0.017446)(1.36, 
0.0179679)(1.41, 0.0184911)(1.46, 0.0190161)(1.51, 0.0195434)(1.56, 
0.0200736)(1.61, 0.0206071)(1.66, 0.0211445)(1.71, 0.0216863)(1.76, 
0.0222328)(1.81, 0.0227847)(1.86, 0.0233423)(1.91, 0.0239062)(1.96, 
0.0244768)(2.01, 0.0250547)(2.06, 0.0256405)(2.11, 0.0262345)(2.16, 
0.0268375)(2.21, 0.02745)(2.26, 0.0280726)(2.31, 0.028706)(2.36, 
0.0293508)(2.41, 0.0300078)(2.46, 0.0306778)(2.51, 0.0313616)(2.56, 
0.03206)(2.61, 0.0327742)(2.66, 0.033505)(2.71, 0.0342537)(2.76, 
0.0350215)(2.81, 0.0358097)(2.86, 0.0366199)(2.91, 0.0374537)(2.96, 
0.0383129)(3.01, 0.0391998)(3.06, 0.0401165)(3.11, 0.0410657)(3.16, 
0.0420505)(3.21, 0.0430743)(3.26, 0.0441412)(3.31, 0.0452559)(3.36, 
0.046424)(3.41, 0.0476523)(3.46, 0.0489489)(3.51, 0.0503241)(3.56, 
0.0517906)(3.61, 0.053365)(3.66, 0.0550696)(3.71, 0.056935)(3.76, 
0.0590061)(3.81, 0.0613529)(3.86, 0.0640958)(3.91, 0.0674801)(3.96, 
0.0722015)
};

\addlegendentry{$\frac{M}{N} = 3.5$}    
\addplot[name path = tf, green, smooth, mark= none] 
coordinates{
(1.01, 0.0626561)(1.06, 0.0643398)(1.11, 0.065726)(1.16, 
0.0670953)(1.21, 0.0684503)(1.26, 0.069793)(1.31, 0.0711254)(1.36, 
0.0724493)(1.41, 0.0737663)(1.46, 0.0750779)(1.51, 0.0763855)(1.56, 
0.0776904)(1.61, 0.0789939)(1.66, 0.0802971)(1.71, 0.0816011)(1.76, 
0.082907)(1.81, 0.0842159)(1.86, 0.0855288)(1.91, 0.0868467)(1.96, 
0.0881707)(2.01, 0.0895016)(2.06, 0.0908406)(2.11, 0.0921886)(2.16, 
0.0935466)(2.21, 0.0949157)(2.26, 0.096297)(2.31, 0.0976916)(2.36, 
0.0991006)(2.41, 0.100525)(2.46, 0.101967)(2.51, 0.103427)(2.56,
0.104906)(2.61, 0.106406)(2.66, 0.10793)(2.71, 0.109478)(2.76, 
0.111052)(2.81, 0.112654)(2.86, 0.114288)(2.91, 0.115954)(2.96, 
0.117656)(3.01, 0.119397)(3.06, 0.121181)(3.11, 0.12301)(3.16, 
0.12489)(3.21, 0.126825)(3.26, 0.128821)(3.31, 0.130885)(3.36, 
0.133024)(3.41, 0.135249)(3.46, 0.13757)(3.51, 0.140001)(3.56, 
0.142562)(3.61, 0.145273)(3.66, 0.148167)(3.71, 0.151285)(3.76, 
0.154688)(3.81, 0.158473)(3.86, 0.162803)(3.91, 0.16801)(3.96, 
0.175036)
};

\addlegendentry{$\frac{M}{N} = 2$}    
\addplot[name path = t,pink, smooth, mark= none] 
coordinates{
(1.01, 0.432057)(1.06, 0.432329)(1.11, 0.430985)(1.16, 
0.429903)(1.21, 0.429052)(1.26, 0.428403)(1.31, 0.427934)(1.36, 
0.427626)(1.41, 0.42746)(1.46, 0.427424)(1.51, 0.427503)(1.56, 
0.427689)(1.61, 0.42797)(1.66, 0.428339)(1.71, 0.428789)(1.76, 
0.429313)(1.81, 0.429905)(1.86, 0.430561)(1.91, 0.431276)(1.96, 
0.432046)(2.01, 0.432867)(2.06, 0.433738)(2.11, 0.434653)(2.16, 
0.435613)(2.21, 0.436613)(2.26, 0.437652)(2.31, 0.438729)(2.36, 
0.439841)(2.41, 0.440988)(2.46, 0.442169)(2.51, 0.443381)(2.56, 
0.444625)(2.61, 0.445901)(2.66, 0.447206)(2.71, 0.448541)(2.76, 
0.449906)(2.81, 0.4513)(2.86, 0.452723)(2.91, 0.454176)(2.96, 
0.455659)(3.01, 0.457172)(3.06, 0.458716)(3.11, 0.460291)(3.16, 
0.461899)(3.21, 0.463542)(3.26, 0.46522)(3.31, 0.466935)(3.36, 
0.468691)(3.41, 0.47049)(3.46, 0.472336)(3.51, 0.474233)(3.56, 
0.476188)(3.61, 0.478208)(3.66, 0.480303)(3.71, 0.482489)(3.76, 
0.484785)(3.81, 0.487223)(3.86, 0.489858)(3.91, 0.4928)(3.96, 
0.496356)
};

\addplot[draw=none,name path = B] {0};     
    \addplot+[pink, opacity = .15] fill between[of=t and B,soft clip={domain=1:4}];
    \addplot+[green, opacity = .15] fill between[of=tf and B,soft clip={domain=1:4}];
    \addplot+[blue, opacity = .15] fill between[of=f and B,soft clip={domain=1:4}];
  
\end{axis}
\end{tikzpicture}
\caption{ }
\label{fig:TE}
\end{figure}

\subsection{$(1 ,3)^D$ channel contributions}
\label{sec13}
$(1,3)^D$ has a dimension  $4\geq 2\Delta_{(1,3)}\geq 2$ for $1\leq Q\leq 4$, decreasing with $Q$. Despite this relatively high dimension, the term
\begin{subequations}
\begin{align}
\mathrm{Leading:}&
\left|w\right|^{-4\Delta_{(0,\frac12)}}\left(\frac{|w|}{N}\right)^{2\Delta_{(1,3)}}\;c_{(1,3)}\\
\mathrm{Sub-leading:}& O\left(\frac{1}{N^{2\Delta_{(1,3)}+2}}\right)
\end{align}
\end{subequations}
where:
\begin{align}
\label{c13}
c_{(1,3)}=\left(2\pi\right)^{2\Delta_{(1,3)}}\; C^{(1,3)^D}_{(0,\frac12),(0,\frac12)}\left< V_{(1,3)^D} (0)\right>_{\tau}
\end{align}
provides a  visible contribution when  $Q\geq 3$, see next Section \ref{further}. 

\noindent We consider then  $\left< V_{(1,3)^D}\right>_{\tau}$. Differently from the case of the energy field, the fusion rule imposed by the  vanishing of the $(1,3)^D$ null state:
\begin{equation}
V_{\Delta_{(1,3)}} \times V_{\Delta_{(r,s)}} \to  V_{\Delta_{(r,s+2)}}\oplus V_{\Delta_{(r,s)}} \oplus  V_{\Delta_{(r,s-2)}},
\end{equation}
does not fix the representations contributing to its one-point function, since  the fusion  $V_{\Delta_{(1,3)}} \times V_{\Delta_{(r,s)}} \to  V_{\Delta_{(r,s)}}$ is allowed for all $r,s$. This can be seen also from the fact that the structure constant $C^{(1,3)^D}_{(\Delta),(\Delta)}$ is different from zero for any $\Delta$ and $c$ \cite{rib14}. Parametrising $\Delta$ as in (\ref{paradelta}), one has, for three diagonal (spinless) fields \cite{dofa_npb85}:

\begin{equation}\label{13sc}
C^{(1,3)^D}_{(r,s)^D,(r,s)^D} = \sqrt{\frac{\gamma^3(\frac{1}{\beta^2})\gamma(2-\frac{2}{\beta^2})\gamma(2-\frac{3}{\beta^2})}{\gamma(\frac{2}{\beta^2})}\frac{\gamma^2(r+\frac{1-s}{\beta^2})}{\gamma^2(1+r-\frac{1+s}{\beta^2})}}
\end{equation}
The above value of the structure constant can be derived either from the vanishing of the third level null state of $(1,3)^D$ or from a Coulomb gas integral, as the three vertex fields satisfy the charge neutrality condition. One can expect on solid grounds that $C^{(1,3)^D}_{(r,s)^D,(r,s)^D}$ describes certain three-point correlation functions in the $Q$-state Potts model. In \cite{ei15} for instance, the structure constant $C^{(1,3)^D}_{(1,0),(1,0)}$ has been checked to correspond to the scaling limit of certain lattice transfer matrix amplitudes. In the case of two non-diagonal fields, $C^{(1,3)^D}_{(r,s),(r,s)}$ has been shown in \cite{ei15,mr17} to be given by  $C^{(1,3)^D}_{(r,s),(r,s)}=\sqrt{C^{(1,3)^D}_{(r,s)^D,(r,s)^D}C^{(1,3)^D}_{(r,-s)^D,(r,-s)^D}}$. 

\noindent One can expect that all the states $X$ in the Potts spectrum  (\ref{sppotts}), such that $C^{(1,3)^D}_{X,X}\neq 0$ contribute to $\left< V_{(1,3)^D}\right>_{\tau}$. However, one has to pay special attention, in particular when using truncations in the $s-$channel spectrum: there can be highly non-trivial cancellations between states. This is known to be the case when the central charge takes rational values, and a finite number of states in the spectrum closes under OPE (see Section 5 of \cite{sv13} and references therein).

\noindent We obtain:
\begin{multline}
\left< V_{(1,3)^D}\right>_{\tau} = \frac{1}{Z_Q}\Big((Q-1)C^{(1,3)^D}_{(0,\frac12),(0,\frac12)}\left|\mathcal{F}^{\Delta_{(0,\frac12)}}_{\Delta_{(1,3)}}(q)\right|^2+C^{(1,3)^D}_{(1,2)^D,(1,2)^D}\left|\mathcal{F}^{\Delta_{(1,2)}}_{\Delta_{(1,3)}}(q)\right|^2\\
+(Q-1)C^{(1,3)^D}_{(0,\frac32),(0,\frac32)}\left|\mathcal{F}^{\Delta_{(0,\frac32)}}_{\Delta_{(1,3)}}(q)\right|^2+\frac{Q(Q-3)}{2} C^{(1,3)^D}_{(2,0),(2,0)}\left|\mathcal{F}^{\Delta_{(2,0)}}_{\Delta_{(1,3)}}(q)\right|^2+\cdots\Big),
\end{multline}
where $\cdots$ indicates next sub-leading contributions. In the above formula, the $Q$ dependent prefactors come again from the multiplicity of the states propagating in the torus. In the following figure, the value $c_{(1,3)}$ in (\ref{c13}) for $M=N$ is plotted as a function of $Q$ in the region of $Q$  where the comparison with Monte Carlo results is possible:
\begin{figure}[H]
\centering
\begin{tikzpicture}
\begin{axis}[
	legend cell align=center,
	xlabel={$Q$},
	ylabel={$c_{(1,3)}$},
	xmin=2.7,
  xmax=3.5, ymin = 0.02, ymax = 0.12
	]
	]

\addplot[red, smooth, mark= none] 
coordinates{
(2.7, 0.0508041)(2.725, 0.0523696)(2.75, 0.0539287)(2.775, 
0.0554826)(2.8, 0.0570319)(2.825, 0.0585777)(2.85, 0.0601208)(2.875, 
0.061662)(2.9, 0.0632021)(2.925, 0.064742)(2.95, 0.0662825)(2.975, 
0.0678243)(3., 0.0693684)(3.025, 0.0709154)(3.05, 0.0724662)(3.075, 
0.0740216)(3.1, 0.0755824)(3.125, 0.0771496)(3.15, 0.078724)(3.175, 
0.0803064)(3.2, 0.0818978)(3.225, 0.0834993)(3.25, 0.0851117)(3.275, 
0.0867363)(3.3, 0.088374)(3.325, 0.0900262)(3.35, 0.0916941)(3.375, 
0.0933791)(3.4, 0.0950827)(3.425, 0.0968065)(3.45, 0.0985523)(3.475, 
0.100322)(3.5, 0.102118)
};
\draw [<-] (2.75,0.055)--(2.75,0.065);\node [above]at (2.75,0.065) {$2.75$};
\draw [<-] (3,0.07)--(3,0.08);\node [above]at (3,0.08) {$3$};
\draw [<-] (3.25,0.086)--(3.25,0.096);\node [above]at (3.25,0.096) {$3.25$};
\draw [<-] (3.41421,0.097)--(3.41421,0.107);\node [above]at (3.41421,0.107) {$2+\sqrt{2}$};
\end{axis}

\end{tikzpicture}
\caption{ }
\label{fig:Q3}
\end{figure}
In the above figure we tagged the values of $Q$ at which Monte Carlo data have been taken. 

\noindent At $Q=3$, only three channels ($\mathcal{S}_{\mathrm{int}}=\{(0,\frac12), (1,2)^D, (1,3)^D\}$) contribute to $\left< V^D_{(1,3)} (0)\right>_{\tau}$, so we expect that, for $Q\sim 3$, $\{(0,\frac12), (1,2)^D\}$ produce the main contributions, while all others are suppressed by some power of $Q-3$.


\section{Monte Carlo simulation and CFT comparisons}
\label{MCsec}
\subsection{General results for the two-point correlation functions}

We collected data on square lattices of size $N\times N$ with periodic boundary conditions on both directions, thus having the topology of a torus (\ref{toruspar}) with $M=N$ (for $M\neq N$ see next subsection).  We considered  various linear sizes $N$ up to $N=8192$.
The probability (\ref{2pointdef}) is computed by considering the lattice points $(x,y)$ and $(x+r,y )$ or $(x,y)$ and $(x,y+r)$ and next averaging over $x$ and $y$. We took data for  $Q=1 + n/4$ for $n=1,\cdots 9$ and $Q=2+2 \cos{\frac{3 \pi}{5}}, 2+\sqrt{2}$. 
For each value of $Q$, we averaged over $10^6$ independent samples generated with the Chayes-Machta Algorithm \cite{chamach98,deng07}. This algorithm is a generalisation of the Swendsen-Wang algorithm for non integer values of $Q$. 

\begin{figure}[h]
\begin{center}
\epsfxsize=320pt\epsfysize=240pt{\epsffile{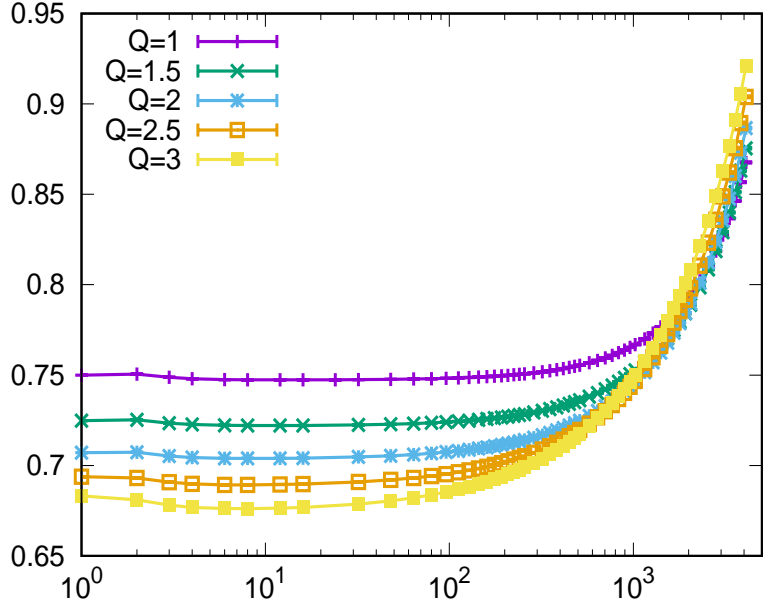}}
\end{center}
\caption{Rescaled two-point connectivity for the $Q$ Potts models at $N=8192$ for various values of $Q$ 
as shown in the caption.}
\label{FigPC2}
\end{figure}

In Fig.~\ref{FigPC2}, we present the rescaled correlation function
$r^{4\Delta_{(0,\frac12)}} p_{12}(r)$ as a function of $r$ for various 
values of $Q$ as shown in the caption. While we observe a plateau for a value 
$\simeq 0.7$, we also see that there exist strong deviations for large $r$. 
This is due to the fact that we work on a torus, thus we expect topological corrections 
which have a maximum at $r=\frac{N}{2}$.  We also need to take into account the
small size corrections which, as can be observed in Fig.~\ref{FigPC2},
will be present only for small sizes up to $r \simeq 10$. 
A general form of fit for the rescaled function $r^{4\Delta_{(0,\frac12)}} p_{12}(r)$ is given by~:
\beq
f(N,r)=c_0 \left(1+\sum_{j \geq 1} c_j \left( \frac{r}{N}\right)^{d_j}\right) \left(1+  g_1\; r^{-g_2 } \right) \; .\label{FF}
\eeq
The above form of fit is factorised into three 
factors. The first factor $c_0$ is the non-universal normalisation of the lattice two-point functions. The second part, with parameters $c_j$, ($j\geq1$) encodes the torus corrections: $d_j$ and $c_j$ are the universal quantities to be compared respectively to the dimensions and the factors computed in the previous sections using CFT, see (\ref{Qgen}). 
The third factor takes into account the small size corrections. In the case of the Ising model, an exact computation shows that 
this correction is described by this form with $g_2=2$ and a small coefficient $g_1 = \frac{1}{64}$ \cite{mcCoy76}.

\begin{figure}[h]
\begin{center}
\epsfxsize=320pt\epsfysize=240pt{\epsffile{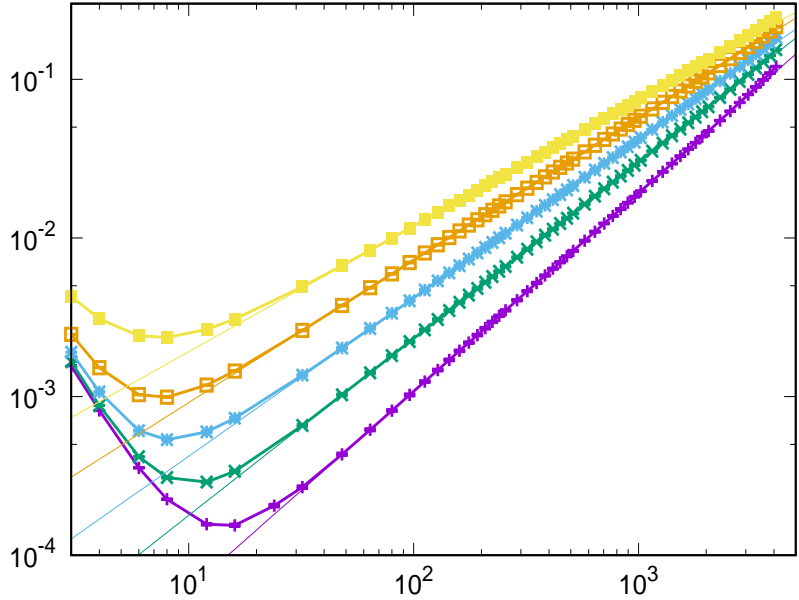}}
\end{center}
\caption{Same data as in Fig.~\ref{FigPC2} with the subtraction of the plateau. 
The thin lines corresponds to best fit as discussed in the text.}
\label{FigQ}
\end{figure}

\noindent A first numerical result is that the dominant topological correction is of the form $\left(\frac{r}{N}\right)^{2\Delta_{(1,2)}}$, i.e. $d_1\sim 2\Delta_{(1,2)}$. In Fig.~\ref{FigQ}, we show the behaviour of $r^{4\Delta_{(0,\frac12)}} p_{12}(r) - c_0 $ with $c_0$ the constant part 
corresponding to the value of the plateau and this for various values of $Q=1, \cdots 3$ as shown in the caption and for $N=8192$. 
We observe that the correction is a power of $r$. We do a fit in the range $r\in [50-200]$ 
obtaining the powers $d_1=\{1.251, 1.115, 0.997, 0.898, 0.793\}$ for $Q \in [1,3]$, which are very close to the 
corresponding set of values of $2\Delta_{(1,2)}=\{1.25, 1.1776, 1, 0.8982,0.8\}$. 
The best fit is also shown in Fig.~\ref{FigQ} as thin lines. 
Note that these fits agree with the numerical data also for much larger distances, $r>200$. 
In the case of $Q=2$, the exact result for the two-point function \cite{FraSaZu87} is~:
\beq 
Q=2,\quad r^{2} \left<V_{(0,\frac12)}(r)V_{(0,\frac12)}(0) \right>_{\tau=i}=1+0.488863 \;\frac{r}{N} + 0.211556 \;\left(\frac{r}{N}\right)^4  + \cdots.
\eeq 
This explains that the leading correction alone gives already a very good fit as shown in Fig.~\ref{FigQ}. We observe that this is also true for other values of $Q$, in agreement with our results (\ref{Qgen}) for $N=M$.

In Tab.~\ref{Table2pt}, we give the numerical results for $c_0$ and $c_1$ obtained with a fit while keeping only 
the leading topological correction and fixing $d_1=2\Delta_{(1,2)}$. The fit is done with numerical data $r\in [6,2048]$. 
With this range of data, we obtain a good fit (measured with the goodness of fit) for each value of $Q$.
The numerical errors on $c_0$ and $c_1$ are indicated in the table either in parenthesis or smaller than one last digit. 
 These fits also take into account small distance corrections. We obtained $g_1 \simeq 0.02$ and $g_2 \simeq 2$ 
 for not too large values of $Q$. Further details on these fits are found in \cite{prs19}. 
\begin{table}[H]
\centering
\begin{tabular}{ |c  |  c | c | c | c |  c | c | } 
\hline
   $Q$   &  $c_0$ &  $c_1 $  & $c_{(1,2)}$  \\
      \hline
   1        &    0.74719  & 0.356  & 0.35707 \\
   1.25   &     0.73323 & 0.392  & 0.393023  \\
$ 2 + \cos{\frac{3 \pi}{5}}$&  0.72693  & 0.414  & 0.411442 \\
   1.5     &     0.72178 & 0.4343   & 0.427244    \\
   1.75   &    0.71199 & 0.459  & 0.458989    \\
   2.0     &   0.70337  &  0.488    & 0.488863    \\
   2.25    &   0.69556 & 0.518 & 0.517293 \\
   2.5      &    0.68827 & 0.551 & 0.544607 \\
   2.75    &   0.68113 & 0.578  & 0.571079  \\
   3.0     &     0.67376 (2) &  0.599 & 0.596962 \\
   3.25   &   0.66555 (5) & 0.627  & 0.622532   \\
  $2+\sqrt{2}$  &  0.65902 (7) & 0.642 & 0.639326 \\
  \hline
\end{tabular}
\caption{$c_0$ and $c_1$ from a fit of the numerical data to the form (\ref{FF}). The last column contains the analytical determination in (\ref{ener})}
\label{Table2pt}
\end{table}
In Table~\ref{Table2pt}, we also show in the last column the values $c_{(1,2)}$ computed in Section \ref{energycontr}.
The agreement is excellent with the numerical value $c_1$, in particular for small values of $Q$. For large values of $Q$, we expect 
that larger corrections have to be taken into account. In order to check the presence of larger corrections 
we can simply attempt a fit to the form (\ref{FF}) while adding a second correction $c_2 (r/N)^{d_2}$. 
We will come back to this point later. 

\subsection{Non-square torus}
\begin{figure}[h]
\begin{center}
\epsfxsize=320pt\epsfysize=240pt{\epsffile{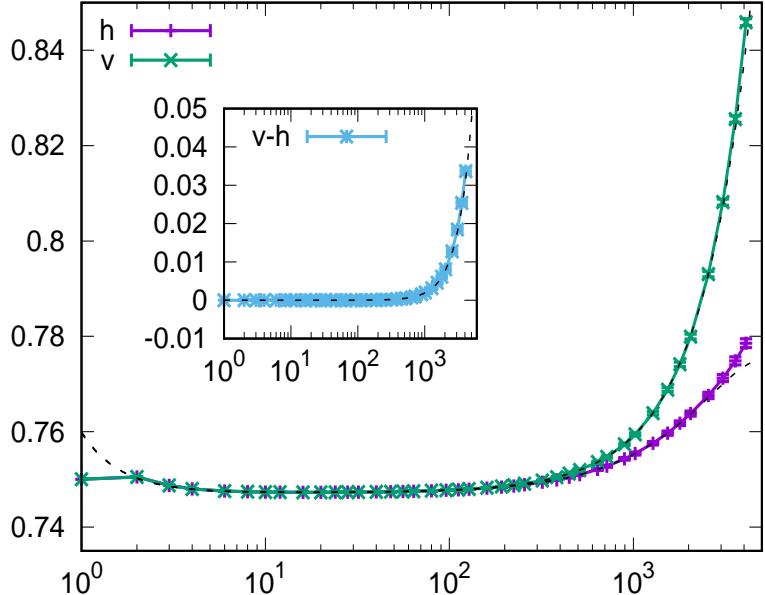}}
\end{center}
\caption{Rescaled two-point connectivity for $Q=1$ and $\frac{M}{N}=2$, at $N=8192$. We show separately the vertical and horizontal connectivities. The inset contains the difference between these two connectivities.}
\label{fig:MN}
\end{figure}

In this section we extend our results to non-square lattices. We checked the agreement between analytical and numerical results for various aspect ratios $\frac{M}{N}$ and for different $Q$'s. Here we present results for $\frac{M}{N}=2$ and $Q=1$, which involves taking a non-trivial limit:  in this regime, the topological correction coming from the stress-energy tensor is non-zero and is given by the finite limit of (\ref{cT}) when $Q\to1$, see Section \ref{idesec}. We consider the correlation measured in the vertical $(v)$ (resp. horizontal $(h)$) directions. The coefficients $c_T^{(v)}$, $c_T^{(h)}$ of $\left(\frac{r}{N}\right)^2$ (resp. $\left(\frac{r}{M}\right)^2$) and $c_{(1,2)}^{(v)}$, $c_{(1,2)}^{(h)}$ of $\left(\frac{r}{N}\right)^{2\Delta_{(1,2)}}$ (resp. $\left(\frac{r}{M}\right)^{2\Delta_{(1,2)}}$) are,
\begin{subequations}
\begin{align*}
&c_T^{(v)} = 2\,c_T(\frac{M}{N}) = 0.175608\quad c_T^{(h)} = 2\left(\frac{M}{N}\right)^{-2}c_T(\frac{N}{M}) = -0.175608\\
&c_{(1,2)}^{(v)} = c_{(1,2)}(\frac{M}{N}) =  0.185569\quad c_{(1,2)}^{(h)} = \left(\frac{M}{N}\right)^{-2\Delta_{(1,2)}}c_{(1,2)}(\frac{N}{M}) = 0.185557.
\end{align*}
\end{subequations}
In Figure \ref{fig:MN} we show the numerical results and the best fits (dashed lines). We obtain 
\begin{align*}
&c_T^{(h)} = -0.165(2)\quad c_T^{(v)} = 0.192(2)\\
&c_{(1,2)}^{(h)} =0.183(1) \quad c_{(1,2)}^{(v)} = 0.180(4).
\end{align*}
The agreement is good. We also show in the inset the difference between vertical and horizontal correlations. This measures directly the contribution of the stress-energy tensor since the contribution of the energy cancels. We obtain $c_T^{(v-h)} = 0.172(1)$.

\subsection{Link with one-point correlation function}
We compare now the value of $c_1$ and the theoretical prediction $c_{(1,2)}$ in (\ref{c12}) to the torus one-point function of the lattice energy field $\left<\varepsilon^{\mathrm{latt}}\right>_{\tau}$. The lattice energy field can be written in terms of the fields in the thermal series  $\mathcal{S}^{D,\mathrm{quot}}_{(1,\mathbb{N^*})}$, see Section (\ref{QPottdef}), giving:  
\beq
\label{ie}
\left<\varepsilon^{\mathrm{latt}}\right>_{\tau} = e_0 + \frac{1}{N^{2\;\Delta_{(1,2)}}} \;e_{1}  +\cdots
\eeq
where  $e_0$ is the usual bulk energy density, associated to the identity $V_{(1,1)^D}$ field, and the sub-leading term $e_{1}$ is related to the energy  $V_{(1,2)^D}$ field:
\begin{equation}
e_{1}= (2\pi)^{2\Delta_{(0,\frac12)}}\;N_\varepsilon^{-1}\; \left<V_{(1,2)^D}\right>_{\tau}.
\end{equation}
In the above formula, $N_\varepsilon$ is the normalisation relating the lattice to the scaling energy field and is computed by determining the energy-energy correlation $ \left<\varepsilon^{\mathrm{latt}}(x)\varepsilon^{\mathrm{latt}}(0)\right>_{\tau}$, in a similar way as we evaluate $c_0$ for the connectivity function.

\noindent In practice, we define the energy operator $\varepsilon^{\mathrm{latt}}(x) $ as the probability that it contains a FK bond. For a given  cluster configuration, $b_o(x)$ is the probability that the site $x=(x_1,x_2)$ is in the same FK cluster as the site $(x_1+1,x_2)$ 
and $b_v(x)$ is the probability that the site $x$ is in the same FK cluster as the site $(x_1,x_2+1)$.
Then the energy operator is defined as 
\begin{equation}
\varepsilon^{\mathrm{latt}}(x) = b_o(x) + b_v(x) -1.
\end{equation}
 This subtraction corresponds to imposing $e_0=0$ in (\ref{ie}). $e_1$ is obtained by measuring $\left<\varepsilon^{\mathrm{latt}}\right>_{\tau} $ and fitting to the form (\ref{ie}). The constant $N_\varepsilon$ is fixed by measuring the two-point energy operator. The measurement for the one-point correlation function have been done on small lattices, up to $N=256$ for the computation 
of $e_1$ and with 100 million samples for each size. We need to use many samples (and then not too big lattices), since $2\Delta_{(1,2)}=O(1)$ and then the deviation from the infinite size is very small. The same is also true for 
$N_\varepsilon$: it is determined from the two-point energy function which decreases very quickly as a function of the distance. The fits were done for distances $r=8-30$ where we ignored small size and topological corrections. 
As a comparison, the measurements for $c_1$ from the two-point correlation function have been done on very large lattices, $N=8192$. 
\begin{figure}[h]
\begin{center}
\epsfxsize=320pt\epsfysize=240pt{\epsffile{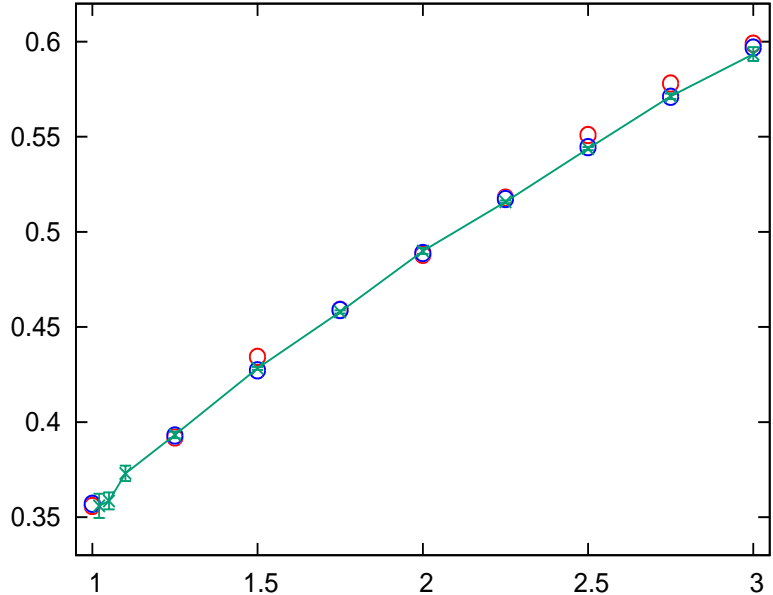}}
\end{center}
\caption{$C^{(1,2)^D}_{(0,\frac12), (0,\frac12)} N_\varepsilon e_1$ vs $Q$ compared to the numerical values $c_1$ shown as red circles and the analytical predictions $c_{(1,2)}$ shown as blue circles.}
\label{Fe}
\end{figure}
In Fig.~\ref{Fe}, we compare the result $C^{(1,2)^D}_{(0,\frac12),(0,\frac12)} \; N_{\varepsilon}\;e_1$ (shown in green) with $c_1$ computed numerically (shown as red circles) and with $c_{(1,2)}$ of (\ref{c12}) (shown as blue circles). 
The agreement between the two measured quantities and the analytical result is very good.

\noindent In the limit $Q\rightarrow 1$, we observe that $C^{(1,2)^D}_{(0,\frac12),(0,\frac12)} \; N_{\varepsilon}\; e_1$ converges to the measured value $c_1$ and $c_{(1,2)}$.
Indeed, we can check numerically that, for $Q \simeq 1$, one has $e_1 \simeq 0.25\;(Q-1)$ while $N_\epsilon \simeq 5.0 \;(Q-1)^{-0.5}$.

\subsection{Further corrections}
\label{further}
We want to check numerically the existence of further  topological corrections. We expect that there exist corrections of order 4 from the descendants of the identity, see Section \ref{idesec} and of order $2 \Delta_{(1,2)}+4$ from the energy descendants, see Section \ref{energycontr}. There exist also the contribution of order $2\Delta_{(1,3)}$, see section \ref{sec13}. In Table \ref{tabdim} we give a comparison of their respective dimensions.
\begin{table}[H]
\center
\begin{tabular}{|c|c|c|c|}
\hline
$Q$	&$2 \Delta_{(1,2)}+4$	&$2\Delta_{(1,3)}$\\\hline
1	&	5.25					&	4			\\
2	&	5					&	3.33			\\
3	&	4.8					&	2.8		 \\\hline
\end{tabular}\caption{ }\label{tabdim}
\end{table}
\noindent For $Q<3$ the coefficient $c_{(1,3)}$ becomes very small (see Figure \ref{fig:Q3} in Section \ref{sec13}), while the dimension $2\Delta_{(1,3)}$ is large and comparable to the dimensions of the descendant fields. Numerically it will then be difficult to distinguish the different contributions for small $Q$'s.

We first compare our numerical data to a fit of the form
\beq
f(N,r)=c_0 \left(1+ c_1  \left( {r\over N }\right)^{d_1} + c_2  \left( {r\over N }\right)^{d_2} \right) (1+  g_1 r^{-g_2 } ) \; ,\label{FF2}
\eeq
Here $d_2$ is an effective dimension which takes into account all possible higher corrections, while we assume the value $d_1 = 2 \Delta_{(1,2)}$ and we take for $c_1=c_{(1,2)}$, see (\ref{c12}). Even so, it is a difficult task since we are left with five parameters. One could try to ignore the small distance corrections
by considering only data at large distances, say $r> r_{min}=50$. This is what we have done for determining the power 
corresponding to the dominant correction. We consider a fit in the range $r_{min} \leq r \leq r_{max}$, with $r_{min}=50$ and $r_{max}=4096$. For the second correction, the fit gives a much less clear image. 
We observe that the second correction is much larger than  $d_1 = 2 \Delta_{(1,2)}$ and its value decreases with $Q$. 
We measure $c_2 \simeq 0.44$ and $d_2 \simeq 5.3$ for $Q=1$ ;  $c_2 \simeq 0.35$ and $d_2 \simeq 4.4$ for $Q=2$; 
$c_2 \simeq 0.29$ and $d_2 \simeq 3.6$ for $Q=3$. We only quote approximate numbers for $c_2$ and $d_2$ since 
they depend on the range $r_{min}$ and $r_{max}$. Still we observe that only for large values of $Q$, \ie\
$Q\simeq 3$, we have a dimension $d_2 < 4$.  This is in agreement with what we expect since it is only for $Q$ close to 3 that the exponent $2\Delta_{(1,3)}$ is smaller than $4$ and $c_{(1,3)}$ becomes non negligible. For smaller values of $Q$, our numerics are not able to give further information.

\noindent For $Q=3$, we can improve by trying a fit 
to the form 
\beq
f(N,r)=c_0 \left(1+ c_1  \left( {r\over N }\right)^{d_1} + c_2  \left( {r\over N }\right)^{d_2} + c_3  \left( {r\over N }\right)^{d_3} \right) \left(1+  g_1 \;r^{-g_2 } \right) \; ,\label{FF3}
\eeq
while imposing the dimensions $d_1=2\Delta_{(1,2)}$, $d_2=2\Delta_{(1,3)}$ and $d_3 = 4$ or $2\Delta_{(1,2)}+4$. In a fit with $r \geq 50$, we obtain a value of $c_2$ in the range $0.05-0.08$ 
(the smallest value is obtained for $d_3 = 4$ and the largest for $d_3=2\Delta_{(1,2)}+4=4.8$), that is comparable with the prediction $c_{(1,3)} \simeq 0.07$ given by (\ref{c13}) for $Q=3$ (see Fig. \ref{fig:Q3}). 

\section{Conclusions}
\label{conclusions}

In this paper we considered  the two-point connectivity $p_{12}$ (\ref{2pointdef}) of the critical $Q-$ random cluster Potts model (\ref{eq:z}) on a torus of parameters (\ref{toruspar}). We focused on the universal corrections to the plane scaling limit of $p_{12}$  originating from the torus topology for general values of $Q\in [1,4]$. Combining CFT techniques with Monte Carlo insights, which suggested the ansatz (\ref{asspect}), we have computed analytically the first dominant corrections to $p_{12}$ in the limit (\ref{limto}). The theoretical results on $p_{12}$, summarised in (\ref{Qgen}), found a very good agreement with Monte Carlo measurements, as shown in Fig. \ref{FigQ} and in  Table \ref{Table2pt}. Moreover, we tested the CFT one-point torus energy function (\ref{e1pt}) against Monte Carlo measurements of the corresponding lattice observable, obtaining again a very good agreement, as shown in Fig.(\ref{Fe}).

Our theoretical results probe non trivial features of the CFT describing the $Q-$state random Potts model, such as the multiplicities of the spectrum (\ref{sppotts}) or the validity of the three-point functions (\ref{ensc}) and (\ref{13sc}) for general values of the central charge. The topological corrections furnish a subtle characterisation of the Potts random clusters which goes beyond the computation of their fractal dimension. As a special case, we obtained the result (\ref{critperco}) that represents a new universal behaviour of critical percolation. 
The study of the torus two-point connectivity represents, together with the plane three-point connectivity \cite{cao2015}, a natural and powerful method to test various conjectures related to critical percolation.

\appendix 

\section{The $s-$ channel expansion of the torus two-point function}
\label{torusdet}
\noindent The Virasoro generators are the modes of the stress-energy tensor. On the plane $z\in \mathbb{C}\bigcup \{\infty\}$, they are defined as:
\begin{equation}
\label{Lndef}
 L^{(z)}_{n} V_{(\Delta,Y)}(z,\bar{z})=\frac{1}{2\pi i} \oint_{\mathcal{C}_z} \;d\;z'\; (z'-z)^{n+1}\;T(z') V_{(\Delta,Y)}(z,\bar{z}),\quad n\in \mathbb{Z},
\end{equation}
Under a  conformal map  $z^\prime = f(z)$, a primary operator transforms:
\begin{align}
\label{primtr}
V_{(\Delta)}(z,\bar{z})=f'(z)^{\Delta}\bar{f}'(\bar{z})^{\bar{\Delta}}\;V_{(\Delta)}(f(z),\bar{f}(\bar{z})),
\end{align}
while the transformation of the Virasoro generators takes the form \cite{EgOo87}\footnote{note that there is a misprint in \cite{EgOo87} for the term $m = n+2$ , as can be checked explicitly in the case $f(z)=z^2$}:
\begin{align}
L_n^{(z)} &= \frac{c}{12}\frac{1}{2\pi i}\oint_z dy(y-z)^{n+1}\{f,y\}+\frac{1}{2\pi i}\oint_z dy \sum_m \frac{L_m^{(f(z))}[f'(y)]^2}{(f(y)-f(z))^{m+2}}(y-z)^{n+1}\nonumber \\
&=\frac{c}{12}\frac{1}{2\pi i}\oint_z dy(y-z)^{n+1}\{f,y\}+[f'(z)]^{-n}L_n^{(f(z))}+\frac{1-n}{2}f''(f')^{-n-2}L_{n+1}^{(f(z))}\nonumber \\
&+\Big(\frac{2-n}{6}f'f'''+\frac{1}{8}(n^2+n-4)(f'')^2\Big)(f')^{-n-4}L_{n+2}^{(f(z))}+\cdots
\label{virtransf}
\end{align}
where $\{f,y\}$ is the Schwarzian derivative.

\noindent To compute torus correlation functions, one needs to know the transformation of (\ref{Lndef})  under the map (\ref{map}). For finite $w$, one obtains for instance:
\begin{subequations}\label{cgen}
\begin{align}
&L_0^{(z)} = L_0^{\mathcal{C},(w)}\nonumber \\
&L_{-1}^{(z)} = z^{-1}\left(\frac{N}{2\pi i}\right)\left(L_{-1}^{\mathcal{C},(w)}-\frac{2\pi i}{N}L_0^{\mathcal{C},(w)}\right) \nonumber \\
&L_{-2}^{(z)} = z^{-2}\left(\frac{N}{2\pi i}\right)^{2}\left(L_{-2}^{\mathcal{C},(w)}-\frac{3}{2}\frac{2\pi i}{N}L_{-1}^{\mathcal{C},(w)}+\frac{13}{12}\left(\frac{2\pi i}{N}\right)^2L_0^{\mathcal{C},(w)}+\left(\frac{2\pi i}{N}\right)^2\frac{c}{24}\right)\nonumber \\
&\cdots \nonumber 
\end{align}
\end{subequations}
The modes  with $L_{n}^{\mathcal{C},(w=\infty)}$, obtained from $L^{(0)}_{n}$ are instead related to contour integrals that are non-contractible on the cylinder. One finds for instance:
\begin{equation}
L_{-n}^{(0)} = L_{-n}^{\mathcal{C},(\infty)}+\frac{c}{24}\delta_{n,0}.
\end{equation}

\noindent Using the above relation, one can easily verify that the one-point torus function of total derivative  $\langle L_{-1}^{\mathcal{C},(0)} V_{(\Delta)}\rangle_{\tau}\propto \langle (L_{-1}^{(1)}+L_0^{(1)})V_{(\Delta)}\rangle$ vanishes,  as can be seen from the vanishing of the matrix elements (\ref{gamma3}):
\begin{multline}
\left<L^{(\infty)}_{Y^\prime}V_{(\Delta^\prime)}L_{-1}^{(1)}V_{(\Delta)}L^{(0)}_{-Y}V_{(\Delta^\prime)}\right>+\left<L^{(\infty)}_{Y^\prime}V_{(\Delta^\prime)}L_{0}^{(1)}V_{(\Delta)}L^{(0)}_{-Y}V_{(\Delta^\prime)}\right> \\
= \left( |Y|-|Y^\prime|-\Delta + \Delta\right)\left<L^{(\infty)}_{Y^\prime}V_{(\Delta^\prime)} V_{(\Delta)}L^{(0)}_{-Y}V_{(\Delta^\prime)}\right> = 0
\end{multline}

\noindent For the two-point function one obtains using (\ref{map}):
%
\begin{equation}
\langle V_{(\Delta_1)}(w_1,\bar{w}_1)V_{(\Delta_2)}(w_2,\bar{w}_2)\rangle_{\tau} = \frac{1}{Z}\text{Tr}_{\mathcal{S}_{\mathrm{int}}} \left(q^{L_{0}^{\mathcal{C},(\infty)}}\bar{q}^{\bar{L}_{0}^{\mathcal{C},(\infty)}}V_{(\Delta_1)}(w_1,\bar{w}_1)V_{(\Delta_2)}(w_2,\bar{w}_2)\right)
\end{equation}
Using (\ref{primtr}) under the map (\ref{map}) \footnote{note that the $i$s drop since the dimensions of our fields satisfy $\Delta-\bar{\Delta} \in2\mathds{Z}$} and the OPE (\ref{OPE}), we find:
\begin{multline}\label{v1v2}
V_{(\Delta_1)}(w_1,\bar{w}_1)V_{(\Delta_2)}(w_2,\bar{w}_2) = \left(\frac{2\pi}{N}\right)^{\Delta_1+\Delta_2}\left(\frac{2\pi}{N}\right)^{\bar{\Delta}_1+\bar{\Delta}_2}z_1^{\Delta_1}z_2^{\Delta_2}\bar{z}_1^{\bar{\Delta}_1}\bar{z}_2^{\bar{\Delta}_2}V_{(\Delta_1)}(z_1,\bar{z}_1)V_{(\Delta_2)}(z_2,\bar{z}_2)\\
=\left(\frac{2\pi}{N}\right)^{\Delta_1+\Delta_2}\left(\frac{2\pi}{N}\right)^{\bar{\Delta}_1+\bar{\Delta}_2}z_1^{\Delta_1}z_2^{\Delta_2}\bar{z}_1^{\bar{\Delta}_1}\bar{z}_2^{\bar{\Delta}_2} \\\times\sum_{(\Delta,Y)}C_{(\Delta)_1,(\Delta)_2}^{(\Delta)}z_{12}^{-\Delta_1-\Delta_2+\Delta+Y}\bar{z}_{12}^{-\bar{\Delta}_1-\bar{\Delta}_2+\bar{\Delta}+\bar{Y}}\tilde{\beta}_{\Delta_1,\Delta_2}^{(\Delta,Y)}\tilde{\beta}_{\bar{\Delta}_1,\bar{\Delta}_2}^{(\bar{\Delta},\bar{Y})}V_{(\Delta,Y)}(z_2,\bar{z}_2)
\end{multline}
where we  made explicit the $z$ dependence of the coefficients (\ref{beta}): $\beta_{\Delta_1,\Delta_2}^{(\Delta,Y)}(z_{12}) = z_{12}^{-\Delta_1-\Delta_2+\Delta+Y}\tilde{\beta}_{\Delta_1,\Delta_2}^{(\Delta,Y)}$. Mapping $V_{(\Delta,Y)}(z_2,\bar{z}_2)$ back to the cylinder: $$V_{(\Delta,Y)}(z_2,\bar{z}_2) = \left(\frac{2\pi i}{N}\right)^{-\Delta-Y}z_2^{-\Delta-Y}\left(-\frac{2\pi i}{N}\right)^{-\bar{\Delta}-\bar{Y}}\bar{z}_2^{-\bar{\Delta}-\bar{Y}}\left(L_{-Y}^{\mathcal{C},w_2}+\cdots\right)\left(\bar{L}_{-\bar{Y}}^{\mathcal{C},\bar{w}_2}+\cdots \right)V^{\mathcal{C}}_{(\Delta)}(w_2,\bar{w}_2)$$ where $\left(L_{-Y}^{\mathcal{C},w_2}+\cdots\right)$ is a linear combination of generators on the cylinder as in relations (\ref{cgen}). 
We can now take the trace, and writing only the holomorphic part we get:
\begin{multline}
\langle V_{(\Delta_1)}(w_1)V_{(\Delta_2)}(w_2)\rangle_\tau=\left(\frac{2\pi }{N}\right)^{\Delta_1+\Delta_2}\left(\frac{z_2}{z_1}\right)^{\Delta_2}\left(1-\frac{z_2}{z_1}\right)^{-\Delta_1-\Delta_2} \\
\times\sum_{(\Delta,Y)\in\mathcal{S}_{\mathrm{int}}}C_{(\Delta)_1,(\Delta)_2}^{(\Delta)}\tilde{\beta}_{\Delta_1,\Delta_2}^{(\Delta,Y)}\left(\frac{2\pi i}{N}\right)^{-\Delta-Y}\left(\frac{z_2}{z_1}\right)^{-\Delta-Y}\left(1-\frac{z_2}{z_1}\right)^{\Delta+Y}  \langle \left(L_{-Y}^{\mathcal{C},w_2}+\cdots\right)V^{\mathcal{C}}_{(\Delta)}(w_2)\rangle_\tau
\end{multline}
Writing $\frac{z_2}{z_1} = e^{-\frac{2\pi i}{N}w_{12}}$ and expanding the exponentials, one has:
\begin{multline}
\langle V_{(\Delta_1)}(w_1)V_{(\Delta_2)}(w_2)\rangle_\tau
=w_{12}^{-\Delta_1-\Delta_2}\sum_{(\Delta)\in\mathcal{S}_{\mathrm{int}}}w_{12}^\Delta  C_{(\Delta)_1,(\Delta)_2}^{(\Delta)}\Bigg(\langle V_{(\Delta)}(w_2)\rangle_\tau\\
+ w_{12}\left\{\frac{2\pi i}{N}\frac{\Delta_1-\Delta_2+\Delta}{2}+\tilde{\beta}_{\Delta_1,\Delta_2}^{(\Delta,1)}(L_{-1}^{\mathcal{C},(w_2)}-\frac{2\pi i}{N}L_0^{\mathcal{C},(w_2)})\right\} \langle V_{(\Delta)}(w_2)\rangle_\tau + O(w_{12}^2)\Bigg)\\
=w_{12}^{-\Delta_1-\Delta_2}\sum_{(\Delta)\in\mathcal{S}_{\mathrm{int}}}w_{12}^\Delta  C_{(\Delta)_1,(\Delta)_2}^{(\Delta)}\bigg(\langle V_{(\Delta)}(w_2)\rangle_\tau\\
+ w_{12}\tilde{\beta}_{\Delta_1,\Delta_2}^{(\Delta,1)} \langle L_{-1}^{\mathcal{C},(w_2)}V_{(\Delta)}(w_2)\rangle_\tau + O(w_{12}^2)\bigg)
\end{multline}
We assume that such cancellations occur at every order in $w_{12}$.  Using the notation (\ref{Lcdef}), we can finally arrive at equation (\ref{2ptorus}). Notice that the coefficients $a^{(\Delta,Y)_\mathrm{int}}_{(\Delta)_1,(\Delta)_2}$ are evaluated using the generators on the plane, while the descendant fields $V^{\mathcal{C}}_{(\Delta,Y)}$ are obtained by acting with the cylinder generators.

\acknowledgments{{We are grateful to Vladimir Dotsenko, 
Benoit Estienne, Christian Hagendorf, Yacine Ikhlef, Jesper Jacobsen, for helpful discussions and we thank in particular  Sylvain Ribault with whom this project started.}

\bibliographystyle{morder6}
\bibliography{letterbib}

\end{document}